# Resonant Ultrasound Spectroscopy: Sensitivity Analysis for Anisotropic Materials with Hexagonal Symmetry

**Christopher L Sevigney**
University of Mississippi
243 Brevard Hall, University MS 38677
clsevign@go.olemiss.edu

**Onome E. Scott-Emuakpor**
Senior Aerospace Engineer
Aerospace Systems Directorate (AFRL/RQTI)
1950 Fifth Street, Bldg. 18D
Wright-Patterson AFB, OH 45433

**Farhad Farzbod**
Assistant Professor, University of Mississippi
201A Carrier Hall, University, MS 38677
farzbod@olemiss.edu

## ABSTRACT

*Resonance ultrasound spectroscopy (RUS) is a non-destructive technique for evaluating elastic and an-elastic material properties. The frequencies of free vibrations for a carefully crafted sample are measured, and material properties can be extracted from this. In one popular application, the determination of monocrystal elasticity, the results are not always reliable. In some cases, the resonant frequencies are insensitive to changes in certain elastic constants or their linear combinations. Previous work has been done to characterize these sensitivity issues in materials with isotropic and cubic symmetry. This work examines the sensitivity of elastic constant measurements by the RUS method for materials with hexagonal symmetry, such as titanium-diboride. We investigate the reliability of RUS data and explore supplemental measurements to obtain an accurate and complete set of elastic constants.*

1. **INTRODUCTION**

Resonance ultrasound spectroscopy (RUS) is an increasingly popular non-destructive method for evaluating materials' properties by careful measurement of a sample's resonant vibrations. Some interesting applications include measuring elasticity in human dentin [1], assessing damage in human bone [2], determining crystallographic orientation [3], detecting microstructural effects of heat-treatment in 3-D printed metals [4], and controlling quality of complex additively manufactured parts [5].

Traditional implementations use piezo-electric transducers to excite and detect resonant frequencies of a lightly clamped sample, attempting to approximate free boundary conditions [6]. Newer techniques include non-contact methods using electromagnetic acoustic transducers [7] or lasers [3, 8-10], allowing for in-situ application to high-temperature test environments [9] and implementations with alternative boundary conditions [11].

To extract material properties, the measured resonant peaks are compared with a theoretically calculated spectrum via an error function, and this error is minimized iteratively by modifying the inputs to the spectral estimation. This regression of the material properties is generally accomplished by the Levenberg Marquardt algorithm [12], but recently more sophisticated approaches have been applied to the regression problem, such as Bayesian inference [13] or multinomial classification [14].

With these techniques, one desires that changes to a material property produce proportional effects on the computed spectrum. However this is not always the case, particularly for the elastic constants of anisotropic materials, where for certain crystallographic symmetries experimentalists have noted the weak effect of certain elastic constants on the resonant spectrum [15] and others have shown explicitly that the spectrum is insensitive to changes of elastic constants in certain linear combinations [16, 17]. There has been various works studying sensitivity using partial derivatives. While it is a valuable data, as it was shown previously [16] there might be linear combinations that affects the resonant frequencies more than the others. So, studying individual partial derivatives does not give the complete picture. Furthermore, if the number of independent elastic constants in an anisotropic material increases, say to 5 and more, then the number of possible combinations, increases drastically. In [16, 17] for example, a material with cubic crystal symmetry is studied. To investigate the sensitivity, one elastic constant is assumed to be fixed, and the other two were varied. Then the resonant frequencies were mapped vs two variable elastic constants. It could be easily observed the effect of change in each elastic constant and/or their linear combination. However, if the number of elastic constants increases, there are couple of obstacles; first, considering variation in all elastic constants, increases exponentially. For example to vary each elastic constant by 20, we have to calculate resonant frequencies for $20^5$ cases. This number, becomes computationally prohibitive if we have 21 elastic constants, requiring $10^{20}$ years on a desktop computer. Second, the number of linear combinations also increases making it difficult to conclude about each linear combination. So, in this paper, we are using Sobol analysis for this purpose; we aim to characterize the reliability of monocrystal elastic properties obtained through the RUS method, focusing specifically on anisotropic materials with hexagonal symmetry such as Titanium Di-Boride (TiB$_2$).

The paper is organized in the following manner. Firstly, the mathematical background for calculating resonant modes of free vibrations is reviewed. Second, Sobol analysis is used to screen the effects of the five independent elastic coefficients on the resonant spectrum. Third, stiffness-frequency relationships are analyzed for a subset of these parameters, identified as being potentially prone to measurement uncertainty. Fourth, the effect of experimental error on the results of the inverse problem is demonstrated directly. Finally, measurement of surface acoustic wave (SAW) propagation is discussed as a supplement to RUS data for accurate and complete determination of elastic properties.

### 1.1. Background of RUS Calculation

This generalized approach to the computation of resonant frequencies for an elastic solid was developed by Visscher [18]. The goal is to form the Lagrangian and find the displacement that causes the Lagrangian to assume its extremum value. By doing so, we obtain a solution to the elastic wave equation. [16]

The Lagrangian for the system is given by Eq. (1.1.1), with kinetic and potential energy terms shown in Eq. (1.1.2). The displacement is assumed to be harmonic with frequency $\omega$; mass density $\rho$, and displacement $u$. The Einstein convention is used, and $i,j,k,l=1,2,3$.

$$L = \frac{1}{2}\int_V \left(\rho\omega^2 u_i^2 - C_{ijkl}\frac{\partial u_i}{\partial x_j}\cdot\frac{\partial u_k}{\partial x_l}\right)dV \quad (1.1.1)$$

$$KE = \frac{1}{2}\rho\omega^2 u_i^2 \qquad PE = \frac{1}{2}C_{ijkl}\frac{\partial u_i}{\partial x_j}\cdot\frac{\partial u_k}{\partial x_l} \quad (1.1.2)$$

The Rayleigh-Ritz method is followed to approximate the displacements with a finite functional basis, $\varphi_q$. The powers of Cartesian coordinates are chosen, where the function label $q=(l,m,n)$ denotes a set of three non-negative integers. This is shown in Eq. (1.1.4) and Eq. (1.1.5).

$$u_i = a_{iq}\varphi_q \quad (1.1.4)$$
$$\varphi_q = x^l y^m z^n \quad q = 1,..,R \quad (1.1.5)$$

The dimension, R, of the basis is constrained by a number N as shown in Eq. (1.1.6), and the dimension is given by Eq. (1.1.7). N=10 is typically reasonable when dealing with the first ~50 eigenmodes.[19]

$$l + m + n \leq N \quad (1.1.6)$$
$$R = \frac{3(N+1)(N+2)(N+3)}{6} \quad (1.1.7)$$

Expanding and rearranging the Lagrangian results in Eq. (1.1.8). The volume integrals can be computed without the coefficients $a_{iq}$, so we bring those terms outside.

$$L = \frac{1}{2}a_{iq}a_{i'q'}\rho\omega^2\int_V \delta_{ii'}\varphi_q(x)\varphi_{q'}(x)dV - \frac{1}{2}a_{iq}a_{kq'}\int_V C_{ijkl}\frac{\partial\varphi_q}{\partial x_j}\cdot\frac{\partial\varphi_{q'}}{\partial x_l}dV \quad (1.1.8)$$

Each volume integral results in a matrix, so we will induce the notation given by Eq.'s (1.1.9-10). Both **E** and **Γ** are symmetric, and **E** is positive definite. Additionally, the term $a_{iq}$ can be written as a column vector **a** with transpose **a$^T$**. The Lagrangian can therefore be re-written in the matrix form of Eq. (1.1.11).

$$E' = \int_V \delta_{ii'} \varphi_q(x) \varphi_{q'}(x) dV \tag{1.1.9}$$

$$\Gamma = \int_V C_{ijkl} \frac{\partial \varphi_q}{\partial x_j} \cdot \frac{\partial \varphi_{q'}}{\partial x_l} dV \tag{1.1.10}$$

$$L = \frac{1}{2}(\rho \omega^2 \mathbf{a}^T \mathbf{E}' \mathbf{a} - \mathbf{a}^T \Gamma \mathbf{a}) \tag{1.1.11}$$

To find the extremum value of the Lagrangian, we take the derivative with respect to **a** and set equal to zero. The result is the general eigenvalue problem of Eq. (1.3.12).

$$\frac{dL}{d\mathbf{a}} = \rho \omega^2 \mathbf{E}' \mathbf{a} - \Gamma \mathbf{a} = 0 \tag{1.1.12}$$

The solution of the eigenvalue problem yields the frequencies of free-oscillation and coefficients for reconstructing the modal patterns.

## 2. SENSITIVITY ANALYSIS

Stiffness measurements based on RUS rely on an inverse procedure in which independent elements of the Voigt matrix are iteratively refined to minimize an error function that compares measured spectra with those generated by the forward problem. For this procedure to be effective, there must be a meaningful response of the spectrum to changes in the parameter space. To discuss the suitability of RUS methods to obtain complete elasticity information, we will analyze the sensitivity of the resonant spectrum to the five independent stiffness elements for crystalline materials with hexagonal structure. The Voigt matrix for this class is given by Eq. (2.0.1).

$$\begin{bmatrix} C_{11} & C_{12} & C_{13} & 0 & 0 & 0 \\ C_{12} & C_{11} & C_{13} & 0 & 0 & 0 \\ C_{13} & C_{13} & C_{33} & 0 & 0 & 0 \\ 0 & 0 & 0 & C_{44} & 0 & 0 \\ 0 & 0 & 0 & 0 & C_{44} & 0 \\ 0 & 0 & 0 & 0 & 0 & C_{66} \end{bmatrix}, \quad C_{66} = \frac{C_{11} - C_{12}}{2} \tag{2.0.1}$$

### 2.1. Global Sensitivity Analysis of Frequencies to Elastic Constants

The Sobol Analysis method is used to quantitatively rank the influence of parameters on the output of very complex multiple input models. It has found its way into various engineering applications in which the number of parameters and their combination becomes relatively large. These applications, among others, include Hydrological models of river basins [20] and state-level energy system models [21]. This technique decomposes a scalar valued function of multiple variables into summands of different dimensions. This is shown in Eq. (2.1.1), where $f_0$ is the expected value of $f(X)$.

$$Y = f(X) := f_0 + \sum_i f_i(x_i) + \sum_{i<j} f_{ij}(x_i, x_j) + \cdots + f_{12\ldots n}(x_1, x_2, \ldots, x_n) \tag{2.1.1}$$

where $X = \{x_1, x_2, \ldots, x_n\}$. The variance of this output is represented by partial variances of the expectation of the output for each variable and set of variable interactions, shown in the following Equations:

$$Var(Y) = \sum_i D_i + \sum_{i<j} D_{ij} + \cdots + D_{1,2,\ldots,n} \tag{2.1.2}$$

$$D_i = Var[E(Y|x_i)] \tag{2.1.3}$$

$$D_{ij} = Var[E(Y|x_i, x_j)] - D_i - D_j \tag{2.1.4}$$

A Sobol index is obtained for each variable and group of variables by computing the ratio of the partial variance to the total variance of the output. The main-effect indices describe the extent to which the variance is reduced when a variable is fixed, that is, the contribution of the variable to the total variance. The total-effect index of a variable is the sum of the indices for main-effect and all interaction effects, indicating its overall impact on the output's variance. These are shown in the following equations:

$$S_i = \frac{D_i}{Var(Y)} \quad (2.1.5)$$

$$S_i^T = S_i + \sum_{i<j} S_{ij} + \cdots + S_{1,2,\ldots,n} \quad (2.1.6)$$

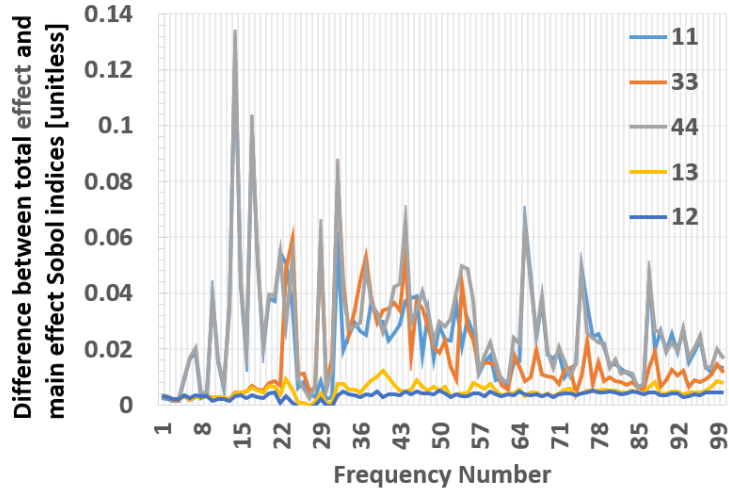

Figure 1: Difference between total effect and main effect Sobol indices for each of the first 100 resonant frequencies of a TiB2 monocrystal with dimensions 2×3×5 mm.

We use Sobol Analysis to quantitatively describe the influence of each of the five independent stiffness coefficients of hexagonal crystals on the variance of each of the first 100 resonant frequencies. For calculations of the resonant spectrum, we consider a theoretical sample of TiB$_2$ with properties from Ledbetter [22]; ρ=4520 kg/m$^3$ , [$C_{11},C_{33},C_{44},C_{12},C_{13}$] = [654,458,262,49,95] GPa, and rectangular parallelepiped geometry with dimensions 2x3x5mm. UQLab's Monte-Carlo based Sobol Analysis toolkit for MATLAB [23, 24] is used to conduct the analysis on each resonant frequency separately. Random sampling of the parameter space is done uniformly in a range of

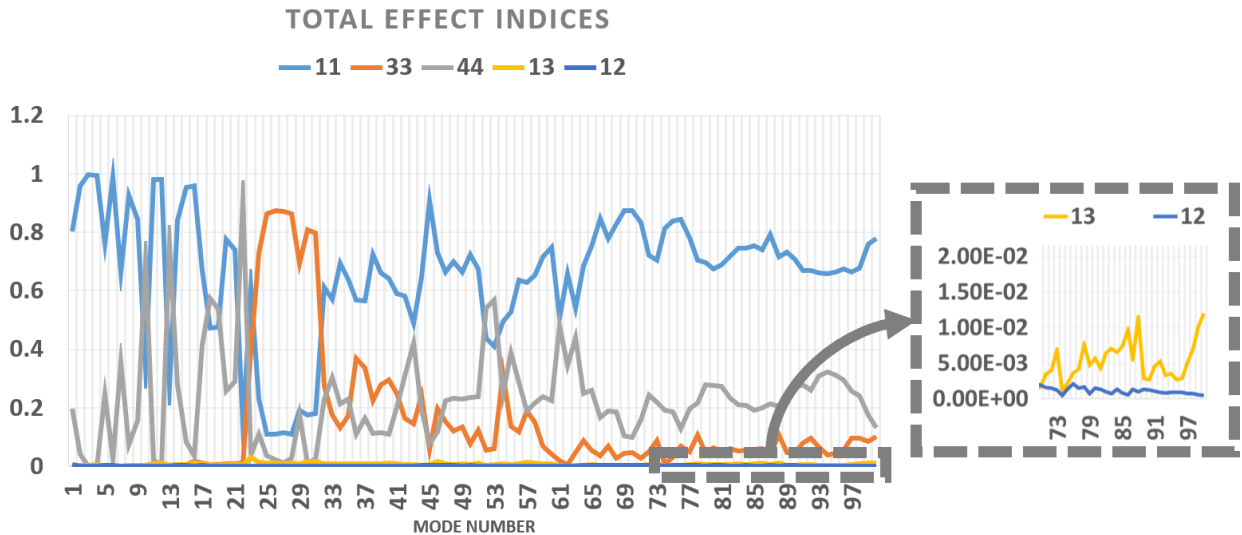

Figure 2: Total Sobol effect indices for each of the first 100 resonant frequencies of a TiB2 monocrystal with dimensions 2x3x5 mm.

+/- 10% from the above stiffness values. The sample size for variance estimation is 100,000 points.

Attention is directed toward the main effect indices, as the difference between total and main effects are small for all frequencies considered and interaction effects are therefore not dominant. This is shown in Figure 1. Of the main effects, it is observed that those stiffness elements lying on the diagonal of the Voigt matrix have some region in the spectrum for which they are main drivers of variance. This is shown in Figure 2. This suggests that changes to these inputs individually produce some significant change to these resonant frequencies, and that small experimental errors in measuring these frequencies should not greatly affect the stiffness values extracted. The small magnitude of the off-diagonal elements suggests a weaker correlation, so a closer look at this relationship is merited.

### 2.2. Frequencies versus Elastic Constants

To better understand the influence of off-diagonal stiffness elements on resonance, the spectrum is calculated for various combinations of $C_{12}$ and $C_{13}$, holding the $C_{11}$, $C_{33}$, and $C_{44}$ constant. Figure 3 shows the ($C_{12}$, $C_{13}$, $f_n$) surface for several mode numbers, *n*. Two observations can be made here: 1) the gradient is small for each of these and 2) the functionality varies for some different mode numbers. For a majority of the modes, however, this variation is small; level sets are generally observed along the approximate lines of $C_{12}+C_{13}$=constant. This is demonstrated by generating an average error function, Eq. (2.2.1), in which the first 100 modes are equally weighted. A 2$^{nd}$ order Taylor series approximation of the error function with respect to these stiffness components is given by Eq. (2.2.2).

$$\chi = \left( \sum_{i=1}^{N} \sqrt{\left(f_i - f_i^{87,42}\right)^2} \Big/ f_i^{87,42} \right) * \frac{100}{N+1} \quad (2.2.1)$$

$$\chi(X + \Delta X) = \chi(X) + \nabla \chi^T \Delta X + \frac{1}{2} \Delta X^T H(X) \Delta X \quad (2.2.2)$$

This hessian term of this expansion produces eigenvalues of the hessian matrix when $\Delta X$ is chosen along an eigenvector, and the expansion is minimized when the eigenvector corresponding to the smallest eigenvalue of the hessian is orthogonal to the gradient. The hessian and gradient terms were approximated numerically [25] at ($C_{12}$,$C_{13}$)=(49,95) GPa, and the eigenvector of the hessian matrix for the smallest eigenvalue is found to have a near-zero dot-product with the gradient when both are unit-normalized (~0.03). This is shown in Figure 4. To further illustrate this point, we observe changes to the resonant spectrum resulting from different $C_{12}$ and $C_{13}$ values along the eigenvector. Let $X = \begin{bmatrix} C_{12} \\ C_{13} \end{bmatrix} = X_0 \pm \Delta X$ represent the stiffness values for comparison, with $X_0 = \begin{bmatrix} 49 \\ 95 \end{bmatrix} GPa$ and $\Delta X = \begin{bmatrix} 15 \\ -15 \end{bmatrix} GPa$. Note that the stiffness step is along $C_{12}+C_{13}$ = constant, and the magnitude of the step is $\frac{\|\Delta X\|_2}{\|X_0\|_2} = 19.85\%$. The relative error between frequencies generated by X and $X_0$ is shown for the first 500 modes in Figure 5. The maximum change to any individual frequency is roughly 1.2% and 1.3%, however most of the frequencies change less than 0.8%. An interesting point is that higher modes see even smaller changes from this step, with most frequencies changing less than 0.5%. The minimal error resulting from the significant steps along this vector means conversely that small errors in capturing the resonant frequencies can significantly impact the value of $C_{12}$ and $C_{13}$ obtained during the inverse problem. This is demonstrated more explicitly in the following section.

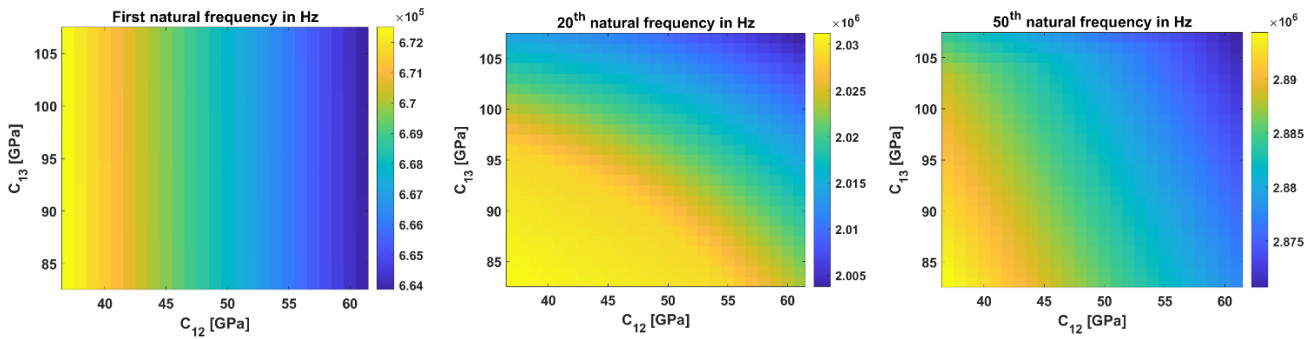

Figure 3: 1st, 20th and 50th resonant frequencies (Hz) for a sample with the dimensions of 2 x 3 x 5 mm and the density of 4520 kg/m3 is calculated for various values of C13 and C12 while (C11, C33, C44 ) was set to (654, 458, 262) GPa.

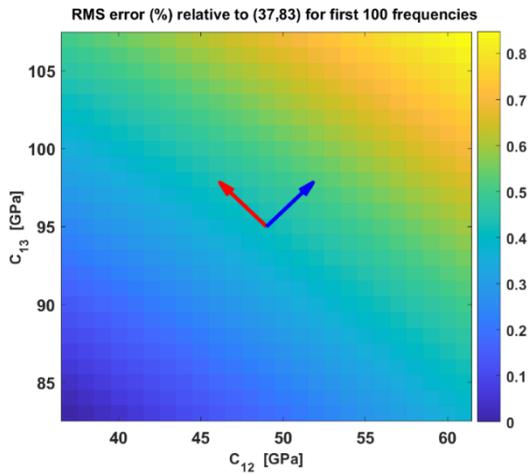

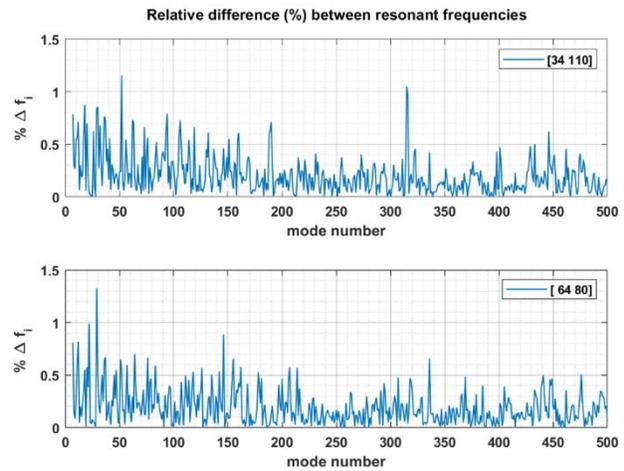

Figure 4: RMS difference (%) between first 100 resonant frequencies of sample with various C12 and C13 versus a sample with C12 and C13 of 37 and 83 GPa. In all cases, the (C11, C33, C44), density and the dimensions are set to be (654, 458, 262) GPa, 4520 kg/m³.

Figure 5: Relative difference (%) between resonant frequencies of two samples with (C12, C13) = (34,110) and (64,80) GPa when they are compared to the ones generated by (C13, C12) = (95,49) GPa. In all cases [C11, C33, C44] are the same and kept at [654 458 262].

### 2.3. Least-Squares Regression from Spectrum with Random Noise

The inverse problem of RUS is generally accomplished by minimizing the sum of squares of the residuals between measured and estimated resonant frequencies via iterative modification of some input parameters, e.g. independent elements of the elastic stiffness. Generally, a subset of twenty to fifty different frequencies is used in this process. Analysis of the forward problem has shown that the values of the elastic stiffness elements can be significantly modified in certain linear combinations while having minimal effect on the average error across many frequencies. In the following section, we observe the effect of small experimental errors in measuring the resonant spectrum on the values of the elastic constants obtained though the regression.

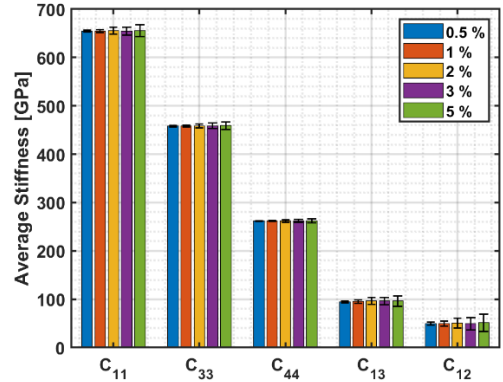

The resonant spectrum for the sample described in Section 2.1 is computed, and the first fifty frequencies are kept. Each of these frequencies is then modified by a random amount, uniformly distributed in a specified range. Then, with a perfect initial guess, the Levenberg-Marquardt algorithm is used to conduct nonlinear least-squares regression on the modified spectrum to obtain a set of theoretical stiffness values. This is done one hundred times for each error envelope. We then compare these theoretical values with the correct values. We consider random errors in the following ranges: ± 0.5%, 1%, 2%, 3%, 5%.

Figure 6: Average and standard deviation (GPa) of stiffness components obtained from Least-Squares regression of first 50 resonant frequencies with random error applied to each frequency in a range of plus/minus the values shown. Procedure is conducted 200 times.

Figure 6 shows the average and standard deviation of the stiffness values obtained from 200 trials of the inverse problem for each of the error envelopes considered. The average value of each component does not change significantly as the maximum frequency error increases;

however, the standard deviation does increase with the maximum frequency error. Figure 7 shows the standard deviation as a percentage of the average stiffness value. The values for the diagonal stiffness components are very consistent for all levels of spectrum error, however regression of the off-diagonal $C_{12}$ and $C_{13}$ components is not robust to these spectrum modifications. Conducting a sufficient number of experiments may yield accurate average values but requires significant time and effort. The following section considers surface acoustic wave measurements as an alternative for improving the reliability of the RUS measurements.

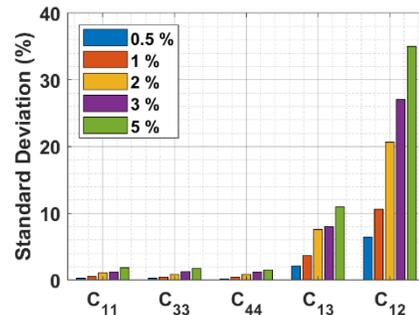

Figure 7: Standard deviations from Fig. 6 shown as a percentage of the corresponding average stiffness values.

### 2.4. Surface Acoustic Wave Velocities and Elastic Constants

Supplementing the RUS method with other experimental techniques has previously been shown as an effective means of improving the quality of elasticity measurements. For example, the addition of pulse-echo data allowed for the acquisition of all 21 independent stiffness elements for an orthotropic material [26]. This method is constrained by sample size requirements, and some hexagonal materials cannot be produced in monocrystal form with sufficient size for this method to be applied in all necessary directions [27]. Here we consider the measurement of surface acoustic wave (SAW) velocity as an alternative. Unlike RUS, SAW is a directional phenomenon; phase speed depends on both the crystallographic plane on which the measurements are taking place and the direction of wave propagation on that plane. Moreover, SAW measurements can be conducted on extremely small surfaces, such as on single grains in polycrystalline samples [28].

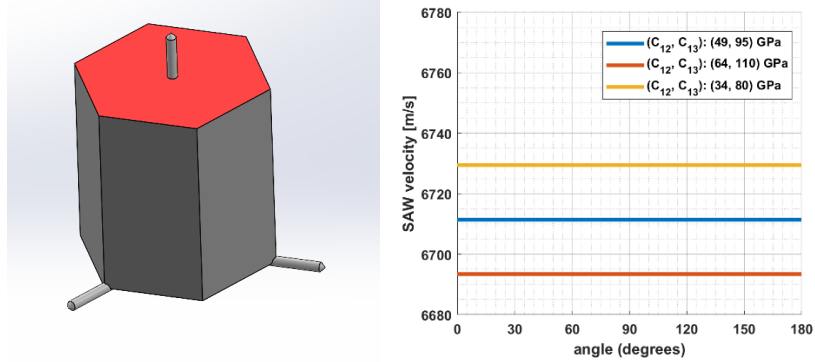

Figure 8: (Left) The [0 0 1] plane of the hexagonal crystal's unit cell. (Right) The directional dependence of SAW velocity for wave vectors on this plane with different $C_{13}$ and $C_{12}$ while ($C_{11}$, $C_{33}$, $C_{44}$) was set to (654, 458, 262) GPa. The angle is measured with respect the <1 0 0> direction.

The mathematical model used for calculation of SAW velocity is based on the work presented by Du [29]. Assuming linear elasticity in a homogeneous and anisotropic half-space, the wave equation is given by Eq. (2.4.1) subject to boundary conditions of Eq. (2.4.2) and the initial conditions Eq. (2.4.3-4). Fourier transform is applied to the spatial (*x,y*) and time (*t*) variables, and the velocities are obtained from the solution in the transformed space.

$$C_{ijkl}\frac{\partial^2 U_l}{\partial x_j \partial x_k} = \rho \frac{\partial^2 U_i}{\partial t} \qquad x_3 \leq 0 \tag{2.4.1}$$

$$\sigma_{i3}|_{z=0} = C_{i3kl}\frac{\partial U_l}{\partial x_k}|_{z=0} = \delta_{i3}\delta(x,y)\delta(t) \qquad i,j,k,l = 1,2,3 \tag{2.4.2}$$

$$U_i(x,y,z,t) = 0 \text{ when } x,y,z,t \to \infty \tag{2.4.3}$$

$$U_i(x,y,z,t) = 0 \text{ for } t < 0 \tag{2.4.4}$$

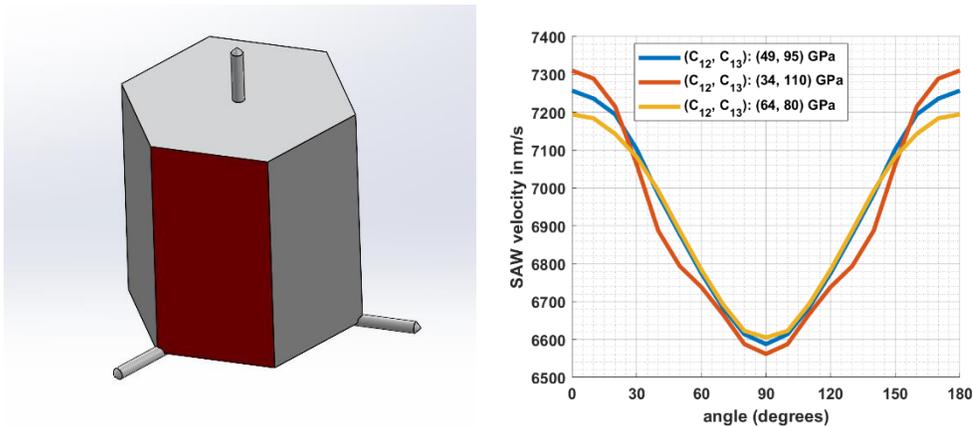

Figure 9: (Left) The [1 0 0] plane of the hexagonal crystal's unit cell. (Right) The directional dependence of SAW velocity for wave vectors on this plane with different $C_{13}$ and $C_{12}$ while ($C_{11}$, $C_{33}$, $C_{44}$) was set to (654, 458, 262) GPa. The angle is measured with respect the <1 0 0> direction.

On the [0 0 1] plane of the hexagonal crystal, illustrated in Figure 8a, the surface wave velocity is independent of the direction; see Figure 8b. This observation is consistent with the plane wave propagation, where longitudinal waves can propagate in any direction on this plane with constant velocity of $c = \sqrt{C_{11}/\rho}$ [30]. However, on other planes with lower symmetry, such as [1 0 0] (Figure 9a), the velocities possess directional dependence. This is shown in Figure 9b for various combinations of $C_{12}$ and $C_{13}$ such that their sum is unchanged. It is evident that the ($C_{12}$,$C_{13}$,velocity) surface has different functionality along these various wave vectors. Figure 10 shows these surfaces for different wave vectors on this plane. By carefully selecting the direction along which to measure the SAW velocity, we can conduct two measurements to extract values for $C_{12}$ and $C_{13}$. The result is depicted in Fig. 11. As can be seen in this figure, there are more error level sets in the combined method, indicating that combining the RUS method with the SAW method, greatly increase the slope of the minimum error resulting in a much better evaluation

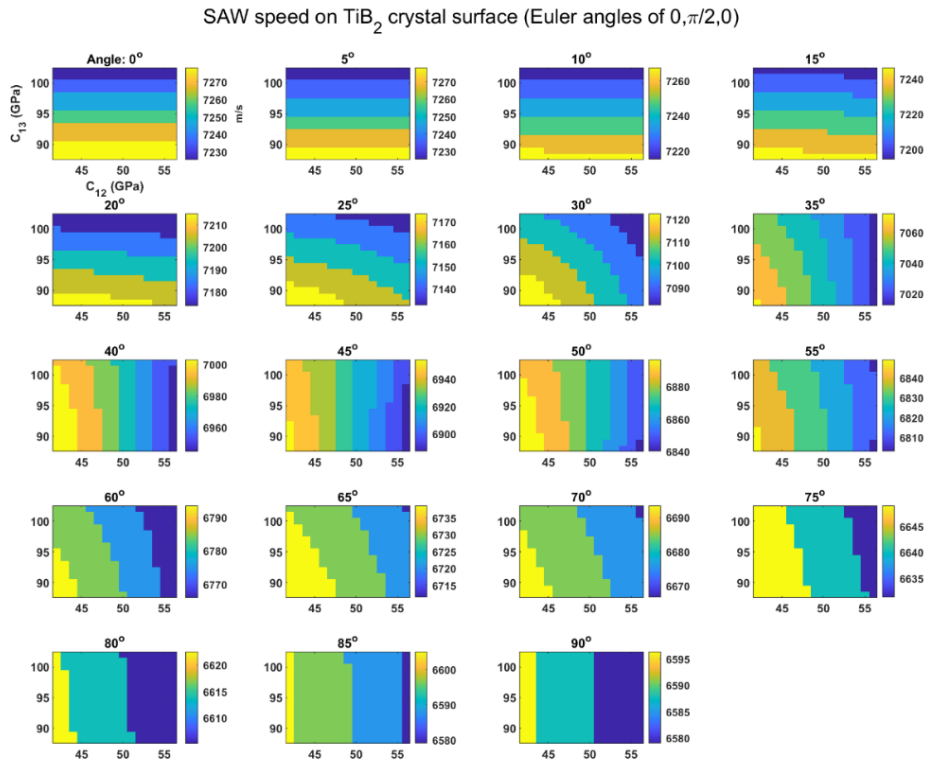

Figure 10: (a) RMS difference (%) between surface acoustic wave velocities on the [1 0 0] plane, sampled from 0 to 180 degrees in 10 degree steps, for a specimen with various $C_{13}$ and $C_{12}$ versus a sample with $C_{13}$ and $C_{12}$ of 95 and 49 GPa. (b) RMS difference (%) between first 100 resonant frequencies. (c) Errors (%) generated by combining the SAW and RUS data with equal weighting.

of the elastic constants.

## 3. CONCLUDING REMARKS

In this article, we investigate the reliability of the RUS method for elasticity measurements of anisotropic materials with hexagonal symmetry. We systematically analyze the influence of each independent element of the Voigt stiffness matrix on the forward problem of RUS. Changes to the off-diagonal elements $C_{12}$ and $C_{13}$ in certain linear combinations are shown to have small effect on the resonant spectrum, suggesting susceptibility to experimental error. This is confirmed by analysis of the inverse problem, where modest errors in the experimental resonant frequencies result in significant variation of these off-diagonal stiffness values. We lastly demonstrate that measurement of surface acoustic waves is an effective supplement to RUS for this material class, enabling more precise determination of the off-diagonal stiffness elements and forming a more reliable characterization of the elasticity.

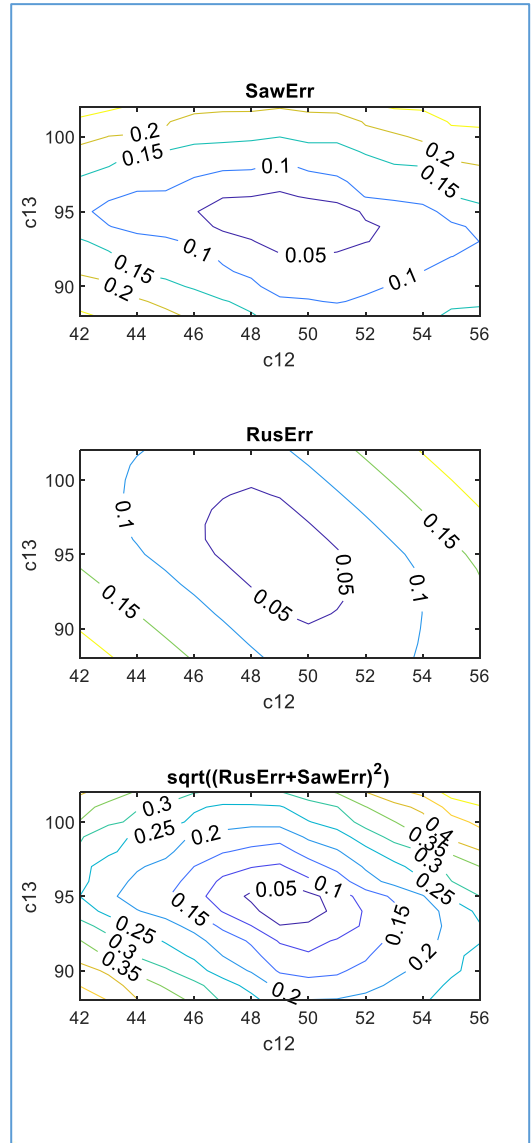

Figure 3: (Top) RMS difference (%) between surface acoustic wave velocities on the [1 0 0] plane, sampled from 0 to 180 degrees in 10 degree steps, for a specimen with various $C_{13}$ and $C_{12}$ versus a sample with $C_{13}$ and $C_{12}$ of 95 and 49 GPa. (Middle) RMS difference (%) between first 100 resonant frequencies. (Bottom) Errors (%) generated by combining the SAW and RUS data with equal weighting.